%% file: root.tex
\documentclass{ifacconf}

\usepackage{natbib}        % required for bibliography
\input{math_commands.tex}

\usepackage{float}
\usepackage{url}
\usepackage{graphicx}
\usepackage{multirow}

\begin{document}
\begin{frontmatter}

\title{Robust Tracking Control with Neural Network Dynamic Models under Input Perturbations
\thanksref{footnoteinfo}} 

\thanks[footnoteinfo]{The first two authors contributed equally. Huixuan Cheng contributed to this work under the affiliation of Carnegie Mellon University as a research intern. This work is in part supported by the National Science Foundation under Grant No. 2144489. The authors would like to thank Xusheng Luo for his help in the revision of the paper.}

%% 修改作者信息，在slack上询问一下
\author[First]{Huixuan Cheng} 
\author[Second]{Hanjiang Hu} 
\author[Second]{Changliu Liu}

\address[First]{Tsinghua University (e-mail: chenghx21@mails.tsinghua.edu.cn).}
\address[Second]{Carnegie Mellon University (e-mail: hanjianghu@cmu.edu,cliu6@andrew.cmu.edu)}

\begin{abstract}                % Abstract of 50--100 words
Robust control problems have significant practical implications since external disturbances can significantly impact the performance of control methods. Existing robust control methods excel at control-affine systems but fail at neural network dynamic models. Developing robust control methods for such systems remains a complex challenge. 
In this paper, we focus on robust tracking methods for neural network dynamic models. We first propose a reachability analysis tool designed for this system and then introduce how to reformulate a robust tracking problem with reachable sets. 
In addition, we prove the existence of a feedback policy that bounds the growth of reachable sets over an infinite horizon. 
The effectiveness of the proposed approach is validated through numerical simulations of the tracking task, where we compare it with a standard tube MPC method.
\end{abstract}

\begin{keyword}
%% chosen from keyword list
Uncertain systems and robust control, Nonlinear control systems, Neural Networks
\end{keyword}

\end{frontmatter}
%%%%%%%%%%%%%%%%%%%%%%%%%%%%%%%%%%%%%%%%%%%%%%%%%%%%%%%%%%%%%%%%%%%%%%%%%%%%%%%%
\section{Introduction}

Neural Network Dynamic Models (NNDMs) have emerged as a powerful alternative for modeling complex systems, leveraging the universal approximation theorem to capture system dynamics through a data-driven approach \citep{liu2023model}. This method simplifies model construction, particularly for systems where deriving accurate analytical models is challenging, time-intensive, or infeasible \citep{nguyen2011model}. Although NNDMs have been successfully applied to various control tasks as described in \cite{hu2024real}, ensuring robust tracking under perturbations in NNDM systems remains an underexplored but crucial area. Given the prevalence of uncertainties and disturbances in real-world applications, robust tracking in NNDM systems is both practically important and technically challenging.

Existing robust control methods, such as robust Model Predictive Control (MPC), are well-established for control-affine systems \citep{liu2014control}. However, these methods are inadequate for addressing the robust tracking problem in NNDMs due to the inherent nonlinearity, limited mathematical interpretability, and black-box nature of NNDMs. Traditional MPC methods, including tube-based MPC \citep{tube1} and its variations \citep{dfmpc1}, are not directly applicable to NNDMs due to these complexities. Other approaches like scenario-tree based MPC \citep{scenariotree1} and system level parameterization \citep{asystemlevel} face scalability issues or are restricted to linear systems.

Our major insight into the robust tracking problem is that we could leverage neural network reachability analysis tools as a function of the control input to analyze the potential tracking error. By analyzing these reachable sets, we can directly minimize tracking errors through optimization of control sequences. This approach draws inspiration from recent advancements in computing approximate reachable sets for polynomial systems under time-varying uncertainties \citep{polynomial1}. While similar methods have been applied to polynomial-nonlinear robust MPC, our approach extends this concept to the more complex domain of NNDMs, addressing the unique challenges posed by neural network dynamics. Our contributions can be summarized as follows.

\begin{itemize}
    \item We formulate a novel multi-step reachability analysis tool designed for NNDMs. 
    \item We propose a robust tracking control method consisting of the reformulation of an optimization problem based on reachability analysis and online optimization of control inputs. %consists offline problem synthesis and online optimization. 
    \item We conducted numerical experiments that demonstrate the efficacy of our method in terms of less control conservativeness and tracking error.
\end{itemize}

\textbf{Notations:} Minkowski sum of two sets $\sA$ and $\sB$ is given
by $\sA\oplus\sB = \{\va+\vb|\va\in\sA,\vb\in\sB\}$. The Pontryagin difference
between two sets $\sA$ and $\sB$ is given by $\sA\ominus\sB = \{\va|\va\oplus\sB\subseteq\sA\}$. The linear mapping between a matrix $\mM$ and a set $\sA$ is given by $\mM\sA = \{\vb| \vb = \mM\va,\va\in\sA\}$.

\section{Problem Formulation and Preliminary}
\subsection{Problem Formulation for NNDM}
\begin{figure*}[!t]
    \centering
    \includegraphics[width=\textwidth]{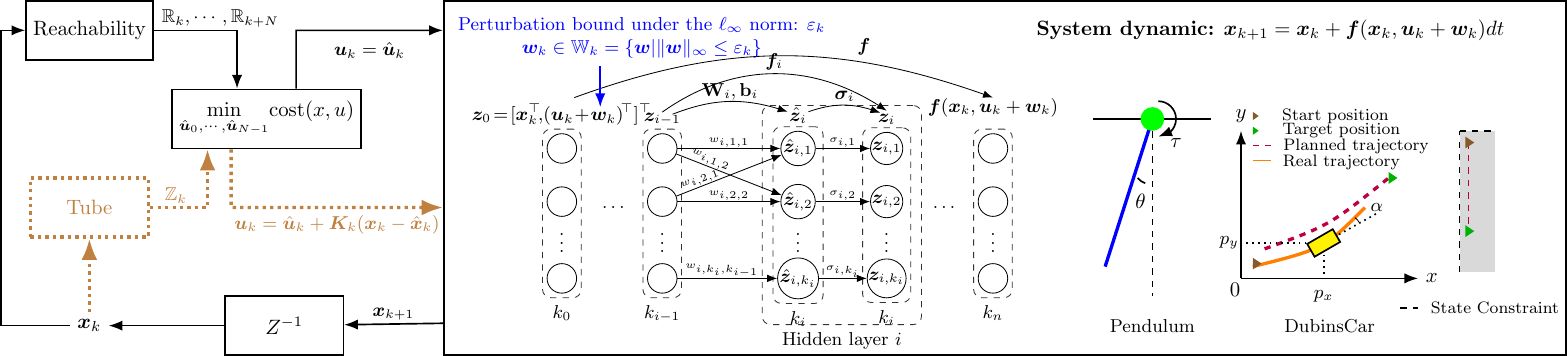}  % 插入 TikZ 代码
    \vspace{-0.6cm}
    \caption{Illustration of the neural network structure, simulation environments, and online control schedule. In neural network, let \(\emW_{ij} \in \sR^{1 \times k_{i-1}}\) be the $j$th row of \(\mW_i\) and \(\evb_{ij}\) the $j$th entry of \(\vb_i\), so \(\hat{z}_{ij} = \emW_{ij} \vz_{i-1} + b_{i,j}\). With ReLU activation, the $j$th entry of \(\vz_i\) is $z_{ij} = \sigma_i(\hat{z}_{ij}) = \max\{0, \hat{z}_{ij}\}$.}
    \label{fig:nn-control-simu}
\end{figure*}

In this work, we consider the discrete-time Neural Network Dynamic Models (NNDM) with bounded perturbations on control inputs defined as follows
\begin{align}
\label{eq:nndm}
\vx_{k+1} = \vx_k + \vf(\vx_k, \vu_k+\vw_k) dt,
\end{align}
where $dt$ is the sampling time interval, $\vf$ is the dynamic model parameterized by a deep neural network, $\vx_{k}\in\sR^{m_x}$ and $\vu_{k}\in\sR^{m_u}$ are state and control input at time $k$, and $\vw_k\in\sR^{m_u}$ is the bounded perturbation applied on $\vu_k$ at time $k$. Assumptions of neural network structure and input perturbation is detailed in Fig. \ref{fig:nn-control-simu}.

System constraints for state and control input are formulated as polytopic sets containing the origin in their interior.
 \begin{align}
 \label{eq:xuconstraint}
 \sX\!=\!\{\vx\in \sR^{m_x}\!\mid\!\mH_x\vx\leq \vh_x\}, \sU\!=\!\{\vu\!\in\!\sR^{m_u}\!|\!\mH_u\vu\!\leq\!\vh_u\}
 \end{align}
 We consider solving infinite robust control problems in a receding horizon fashion by solving $N$ horizon predictive control at every time step $k$ as follows.
% The control problem is formulated as a N-step MPC tracking problem with quadratic costs and linear constraints. %Optimization object for tracking problem at each time step $k$ is defined as one-step MPC with quadratic control cost.
%The overall optimization problem for robust tracking is defined as follows
\begin{prob}
    \label{pb:robust-tracking}
\begin{equation}
    \begin{aligned}
        \text{min}& \sum_{i=k}^{k+N-1}\left\{(\vx_{i}\!-\!\vr_{i})^\top\mQ (\vx_{i}\!-\!\vr_{i})\!+\!\vu_{i}^\top\mR \vu_i\right\}\\
        &+ (\vx_{k+N}-\vr_{k+N})^\top\mQ_f (\vx_{k+N}-\vr_{k+N})\\
        \text { s.t. } &  \vx_{i+1} = \vx_i + \vf(\vx_i, \vu_i+\vw_i) dt, \\
        & \vx_{i} \in \sX, \vx_{k+N}\in\sX_f, \vu_i+\vw_i \in \sU, \vw_i\in \sW_i\\
        & i = k,k+1,\cdots,k+N-1
    \end{aligned}
\end{equation}
where $\mQ$, $\mR$ are positive semi-definite cost matrix, $\mQ_f$ is terminal cost matrix, $N$ is control horizon, $\vu_i$ is control variable, $\vr_{i}$ is reference state, $\sX$ and $\sU$ are the state and action constraint defined in (\ref{eq:xuconstraint}), $\vw_i$ and $\sW_i$ are sampled perturbation and bounded perturbation set respectively. Note that $\vx_k$ is the initial state and is not subject to optimization. \end{prob} 

\subsection{Preliminaries}
\subsubsection{Tube MPC}
%\begin{preliminary}
\label{pl:tubempc}
The Tube MPC methods \citep{tube1} address robust control for nonlinear systems by locally linearizing via Taylor series expansion at each time step. The linearized system at time $k$ is given by $(\mA_k,\mB_k)$.
\begin{align}
\vx_{k+1} = \mA_k\vx_k+\mB_k\vu_k+\mB_k\vw_k\\
\label{eq:tubelinearsystem}
\mA_k = \frac{\partial{\vf}}{\partial{\vx_k}}dt+\mI,\mB_k = \frac{\partial{\vf}}{\partial{\vu_k}}dt
\end{align}
Then,tube MPC separates dynamics with perturbations into a nominal system and a state error system, using feedback control to bound the state error:
\begin{align}
\label{eq:tube-property}
&\text{Nominal system} \quad\hat{\vx}_{k+1} = \mA_k\hat{\vx}_k+\mB_k\hat{\vu}_k\\
&\text{State error system} \ \ve_{k+1}\!= \!(\!\mA_k\!+\!\mB_k\mK_k\!)\ve_k\!+\!\mB_k\vw_k\\
&\text{Feedback control}\quad \vu_k=\hat{\vu}_k\!+\!\mK_k(\vx_k-\hat{\vx}_k)
 \end{align}
 where $\hat{\vx}_k, \hat{\vu}_k$ are nominal state and control input without perturbations, $\ve_k = \vx_k-\hat{\vx}_k$ is defined as state error. The feedback gain $\mK_k$, determined by $\mA_k,\mB_k$, ensures asymptotic stability of the state error system.
 \begin{defn}
 \textbf{Disturbance invariant set} $\sZ_k$ for linearized system $(\mA_k, \mB_k)$ at time $k$ \citep{disturbanceinvariant},
  satisfies 
  \begin{align}
   \label{def:disturbanceinvariantset}
  &\text{for}\ \forall \ve_i\in\sZ_k, \forall\vw_i\in\sW_i, i = 0,1,\cdots\\
  &\ve_{i+1} = (\mA_k + \mB_k\mK_k)\ve_i+ \mB_k\vw_i\in\sZ_k 
  \end{align}
where feedback $\vu_i= \hat{\vu}_i+\mK_k(\vx_i-\hat{\vx}_i)$ is applied to bound the growth of $\sZ_k$. This ensures that for a fixed system $(\mA_k, \mB_k)$, state errors remain bounded within $\sZ_k$ over time despite bounded disturbances. 
 \vspace{-0.2cm}
 \end{defn}
%  \begin{defn}
%  \textbf{Nominal optimization problem} is to solve the tracking problem with the nominal system as dynamic constraints. It can be formulated as:
%   \begin{equation}
%     \begin{aligned}
%         \text{min}& \sum_{i=0}^{N-1}((\hat{\vx}_{i}\!-\!\vr_{i})^\top\mQ (\hat{\vx}_{i}\!-\!\vr_{i})\!+\!\hat{\vu}_{i}^\top\mR \hat{\vu}_i)\\
%         &+ (\hat{\vx}_{N}-\vr_{N})^\top\mQ_f (\hat{\vx}_{N}-\vr_{N})\\
%         \text { s.t. } &  \hat{\vx}_{i+1} = \hat{\vx}_i + \vf(\hat{\vx}_i, \hat{\vu}_i) dt, \\
%         & \hat{\vx}_{i} \in \sX, \hat{\vx}_{N}\in\sX_f, \hat{\vu}_i \in \sU\\
%         & i = 0,1,\cdots,N-1
%     \end{aligned}
% \end{equation}
%  \end{defn}
The disturbance invariant set for the linearized system at time $k$ can be computed as an infinite Minkowski sum: $\sZ_k = \mB_k\sW_k\oplus(\mA_k + \mB_k\mK_k)\mB_k\sW_k\oplus(\mA_k + \mB_k\mK_k)^2\mB_k\sW_k\oplus\cdots$. 
% The robustness of the tracking problem can be guaranteed by incorporating $\sZ_k$ into the constraints of the nominal optimization problem and applying feedback control according to the nominal solutions. The nominal tracking problem at time step $k$ treats non-perturbated system dynamic as the constraint. The non-perturbated variables are denoted as: $\hat{\vx}_{i},\hat{\vu}_{i}$ and the problem can be reformulated as follows
Tracking robustness in Problem \ref{pb:robust-tracking} is ensured by incorporating $\sZ_k$ into the nominal optimization constraints and applying feedback control. The nominal problem at time step $k$ treats nominal dynamics as constraints and is reformulated as:
\begin{prob}
\label{pb:robust-tracking-tube}
\begin{equation*}
    \begin{aligned}
        \text{min}& \sum_{i=k}^{k+N-1}\left\{(\hat{\vx}_{i}\!-\!\vr_{i})^\top\mQ (\hat{\vx}_{i}\!-\!\vr_{i})\!+\!\hat{\vu}_{i}^\top\mR \hat{\vu}_i\right\}\\
        &+ (\hat{\vx}_{k+N}-\vr_{k+N})^\top\mQ_f (\hat{\vx}_{k+N}-\vr_{k+N})\\
        \text { s.t. } &  \hat{\vx}_{i+1} = \hat{\vx}_i + \vf(\hat{\vx}_i, \hat{\vu}_i) dt, \hat{\vx}_k = \vx_k\\
        & \hat{\vx}_{i} \in \sX\ominus\sZ_k, \hat{\vx}_{N}\in\sX_f\ominus\sZ_k, \\
        &\hat{\vu}_i \in \sU\ominus\mK_k\sZ_k\ominus\sW_i, i = k,k+1,\cdots,k+N-1
    \end{aligned}
\end{equation*}
\end{prob}

% After solving the nominal problem for $\hat{\vu}_0,\cdots,\hat{\vu}_{N-1}$, applying feedback control$\vu_k = \hat{\vu}_k+\mK_k(\vx_k-\hat{\vx}_k)$ ensures constraint satisfaction in Problem~\ref{pb:robust-tracking} under disturbances.

% 是否应该加上nominal system 的constraints？
% disturbance invarinat set 加一个定义，可以参考original tube MPC的定义
%\end{preliminary}
\vspace{-0.2cm}
\subsubsection{Reachability for Neural Networks}

\label{pl:verificationpreliminary}
The reachability analysis for NNDMs is mainly based on neural network verification methods \citep{crown2018}, which provide symbolic linear bounds for the output of the neural network given perturbed inputs within a specific range.
\begin{defn}
\textbf{Symbolic linear bound} is a linear inequality involving symbolic variables. For a neural network with input data $[\vx_{k-1}^\top,\vu_{k-1}^\top]^\top$ perturbated within a $\epsilon$-bounded $l_p$-ball, the symbolic linear bound is computed as:
\begin{align}
    \underline{\bm{W}}_{x_k}^x \left[\begin{array}{l}\bm{x}_{k-1} \\ \bm{u}_{k-1}\end{array}\right]\!+\!\underline{\bm{b}}_{x_k}^x \leq
    \bm{x}_{k}\!\leq\!
    \overline{\bm{W}}_{x_k}^x \left[\begin{array}{l}\bm{x}_{k-1} \\ \bm{u}_{k-1}\end{array}\right]\!+\!\overline{\bm{b}}_{x_k}^x
    \label{eq:1layer}
\end{align}
where $\underline{\bm{W}}_{x_k}^x,\overline{\bm{W}}_{x_k}^x\in \sR^{m_x\times(m_x+m_u)},\underline{\bm{b}}_{x_k}^x,\overline{\bm{b}}_{x_k}^x\in\sR^{m_x}$ are the linear weights and biases for the lower and upper bounds respectively, dependent on $[\vx_{k-1}^\top,\vu_{k-1}^\top]^\top$ and $\epsilon$. Specifically, $\vx_{k-1},\vu_{k-1}$ are symbolic variables that can be assigned specific values within the perturbation bounds, ensuring the soundness of inequality. 
\end{defn}
The symbolic linear bounds are primarily derived using interval arithmetic at the hidden layers and linear relaxations at ReLU-activated layers. The calculation of $\underline{\bm{W}}_{x_k}^x, \overline{\bm{W}}_{x_k}^x, \underline{\bm{b}}_{x_k}^x, \overline{\bm{b}}_{x_k}^x$ is propagated from the input layer to the output layer, based on hidden layer parameters, and the selection of linear relaxation bounds for the ReLU units. 
%\end{preliminary}
\subsubsection{Challenges}
Based on the information above, the challenges of solving robust tracking problem with perturbations on control input can be summarized as follows
\begin{itemize}
\item The tube MPC method introduces errors into the disturbance invariant set due to linearizing nonlinear dynamics, often leading to over conservativeness~\citep{tubempcshrink}.
\item Verification methods typically provide bounds for neural networks rather than directly addressing NNDMs, making it non-trivial to obtain tight reachability analysis over the look-ahead horizon.
\end{itemize}

\section{Reachability-based Robust MPC for NNDMs}

%\subsection{Overview}
In this section, we first present how to develop multi-step reachability analysis for NNDMs without linearization of the system dynamics. Next, we introduce the reformulation of the robust tracking problem given the reachable set. Finally, we show the procedure for online execution and how the optimization problem is solved online. 

\subsection{Reachability analysis and robust tracking problem }
We present a $k$-step reachability analysis tool for NNDMs based on the verification method, bridging the gap between neural network verification and reachability computation for NNDMs.

\begin{defn}
\label{def:+-operation}
The operations $(\cdot)_{+}$ and $(\cdot)_{-}$ are defined as $(\cdot)_{+} = ReLU(\cdot)$ and $(\cdot)_{-} = -ReLU(-\cdot)$, where ReLU is applied elementwise.
\end{defn}

\begin{thm}
For a NNDM system satisfying (\ref{eq:nndm}), $k$ is any time step, $\vx_0$ is the initial state, the reachable set of $\bm{x}_k$ can be linearly bounded by $\bm{x}_{0}, \bm{u}_{0}, \cdots, \bm{u}_{k-2}, \bm{u}_{k-1}$ variables. Weights and biases of linear bound w.r.t. these variables can be obtained recursively as follows:
{
\small
\begin{align}
    \underline{\mathbf{W}}_k^x \left[\begin{array}{c} \bm{x}_{0} \\ \bm{u}_{0}\\ \cdots \\ \bm{u}_{k-1}\end{array}\right]+\underline{\mathbf{b}}_k^x \leq \bm{x}_k \leq
    \overline{\mathbf{W}}_k^x \left[\begin{array}{c} \bm{x}_{0} \\ \bm{u}_{0}\\ \cdots \\ \bm{u}_{k-1}\end{array}\right] + \overline{\mathbf{b}}_k^x
    \label{eq:ueq}
\end{align}
}
where
{
\small
$
    \underline{\mathbf{W}}_1^x = \underline{\bm{W}}_{x_k}^x, \overline{\mathbf{W}}_1^x = \overline{\bm{W}}_{x_k}^x;
    \underline{\mathbf{b}}_1^x = \underline{\bm{b}}_{x_k}^x, \overline{\mathbf{b}}_1^x = \overline{\bm{b}}_{x_k}^x
$
\[
\left\{
\begin{array}{l}
\vspace{0.1cm}
\underline{\mathbf{W}}_{i+1}^x\!=\! (\underline{\mathbf{W}}_{i}^x)_{+}\!\left[
\begin{array}{@{}cc@{}}\underline{\bm{W}}_{x_{k-i}}^x &\vzero\\\vzero&\mI_{i\cdot m_u} \end{array}\right]
\!+\!(\underline{\mathbf{W}}_{i}^x)_{-}\!\left[
\begin{array}{@{}cc@{}}\overline{\bm{W}}_{x_{k-i}}^x &\vzero\\\vzero&\mI_{i\cdot m_u} \end{array}\right]\\
\vspace{0.1cm}
\overline{\mathbf{W}}_{i+1}^x\!=\!(\overline{\mathbf{W}}_{i}^x)_{+}\!\left[
\begin{array}{@{}cc@{}}\overline{\bm{W}}_{x_{k-i}}^x &\vzero\\\vzero&\mI_{i\cdot m_u} \end{array}\right]
\!+\!(\overline{\mathbf{W}}_{i}^x)_{-}\!\left[
\begin{array}{@{}cc@{}}\underline{\bm{W}}_{x_{k-i}}^x &\vzero\\\vzero&\mI_{i\cdot m_u} \end{array}\right]\\
\vspace{0.1cm}
\underline{\mathbf{b}}_{i+1}^x\!=\!\underline{\mathbf{b}}_{i}^x
+(\underline{\mathbf{W}}_i^x)_{+}\left[\begin{array}{c}\underline{\bm{b}}_{x_{k-i}}^x\\ \vzero_{i\cdot m_u}\end{array}\right]+(\underline{\mathbf{W}}_i^x)_{-}\left[\begin{array}{c}\overline{\bm{b}}_{x_{k-i}}^x\\ \vzero_{i\cdot m_u}\end{array}\right]\\
\overline{\mathbf{b}}_{i+1}^x\!=\!\overline{\mathbf{b}}_{i}^x
+(\overline{\mathbf{W}}_i^x)_{+}\left[\begin{array}{c}\overline{\bm{b}}_{x_{k-i}}^x\\ \vzero_{i\cdot m_u}\end{array}\right]+(\overline{\mathbf{W}}_i^x)_{-}\left[\begin{array}{c}\underline{\bm{b}}_{x_{k-i}}^x\\ \vzero_{i\cdot m_u}\end{array}\right]
\end{array}
\right.
\]
% \vspace{-0.1cm}
$$\underline{\mathbf{W}}_i^x,\overline{\mathbf{W}}_i^x \in \sR^{m_x \times (m_x+i\cdot m_u)}, \underline{\mathbf{b}}_i^x,\overline{\mathbf{b}}_i^x\in \sR^{m_x},\\
i= 1,2,\cdots,k-1,$$
} 
and 
$\overline{\bm{W}}_{x_{k-i}}^x, \underline{\bm{W}}_{x_{k-i}}^x, \underline{\bm{b}}_{x_{k-i}}^x, \overline{\bm{b}}_{x_{k-i}}^x$ represent the weights and biases for the symbolic linear bound of the state $\vx_{k-i}$, derived using (\ref{eq:1layer}). Similarly, $\underline{\mathbf{W}}_k^x, \underline{\mathbf{b}}_k^x, \overline{\mathbf{W}}_k^x, \overline{\mathbf{b}}_k^x$ define the linear bounds of the reachable set for \(\vx_k\), based on the input sequence region $(\bm{x}_{0}, \bm{u}_{0}, \cdots, \bm{u}_{k-2}, \bm{u}_{k-1})$. 
\label{thm:recursivebound}
\end{thm}
\begin{pf}
We prove (\ref{eq:ueq}) by proving a stronger inequality as follow, where $l = 1,\cdots, k$
\vspace{-0.1cm}
{
\small
\begin{align}
    \underline{\mathbf{W}}_l^x \left[\begin{array}{c} \bm{x}_{k-l} \\ \bm{u}_{k-l}\\ \cdots \\ \bm{u}_{k-1}\end{array}\right]+\underline{\mathbf{b}}_l^x \leq \bm{x}_k \leq
    \overline{\mathbf{W}}_l^x \left[\begin{array}{c} \bm{x}_{k-l} \\ \bm{u}_{k-l}\\ \cdots \\ \bm{u}_{k-1}\end{array}\right] + \overline{\mathbf{b}}_l^x 
    \label{eq:ueqstronger}
\end{align}
}
\vspace{-0.05cm}
We use mathematical induction to prove (\ref{eq:ueqstronger}) holds for $l=1,\cdots,k$, with the case $l = k$ establishing Theorem \ref{thm:recursivebound}. First, when $l=1$, (\ref{eq:ueqstronger}) reduces to a symbolic linear bound in (\ref{eq:1layer}), which holds trivially. Assume (\ref{eq:ueqstronger}) holds for $l=n,(n<k-1)$, where the reachable set of $\vx_k$ can be linearly bounded by $\vx_{k-n},\vu_{k-n},\cdots,\vu_{k-1}$.  We then substitute symbolic bound of $\vx_{k-n}$ into $l=n$ inequality and rearrange it as follows
\vspace{-0.2cm}
{\!
\small
\begin{align*}
    & \left(\!(\underline{\mathbf{W}}_{n}^x)_{+}\!\left[
\begin{array}{@{}cc@{}}\underline{\mathbf{W}}_{x_{k\!-\!n}}^x &\vzero\\\vzero&\mI_{n\cdot m_u} \end{array}\right]
\!+\!(\underline{\mathbf{W}}_{n}^x)_{-}\!\left[
\begin{array}{@{}cc@{}}\overline{\mathbf{W}}_{x_{k\!-\!i}}^x &\vzero\\\vzero&\mI_{n\cdot m_u} 
\end{array}\right]\!\right)\!
\left[\begin{array}{@{}c@{}} {\mathbf{x}}_{k\!-\!n\!-\!1} \\ \mathbf{u}_{k\!-\!n\!-\!1}\\ \cdots \\ \mathbf{u}_{k\!-\!1}\end{array}\right]\\
& +\underline{\mathbf{b}}_{n}^x
+(\underline{\mathbf{W}}_n^x)_{+}\left[\begin{array}{c}\underline{\mathbf{b}}_{x_{k-n}}^x\\ \vzero_{n\cdot m_u}\end{array}\right]+(\underline{\mathbf{W}}_n^x)_{-}\left[\begin{array}{c}\overline{\mathbf{b}}_{x_{k-n}}^x\\ \vzero_{n\cdot m_u}\end{array}\right] \leq \mathbf{x}_k \leq \\
& \left(\!(\overline{\mathbf{W}}_{n}^x)_{+}\!\left[
\begin{array}{@{}cc@{}}\overline{\mathbf{W}}_{x_{k\!-\!n}}^x &\vzero\\\vzero&\mI_{n\cdot m_u} \end{array}\right]
\!+\!(\overline{\mathbf{W}}_{n}^x)_{-}\!\left[
\begin{array}{@{}cc@{}}\underline{\mathbf{W}}_{x_{k\!-\!i}}^x &\vzero\\\vzero&\mI_{n\cdot m_u} 
\end{array}\right]\!\right)\!
\left[\begin{array}{@{}c@{}} {\mathbf{x}}_{k\!-\!n\!-\!1} \\ \mathbf{u}_{k\!-\!n\!-\!1}\\ \cdots \\ \mathbf{u}_{k\!-\!1}\end{array}\right]\\
& +\overline{\mathbf{b}}_{n}^x
+(\overline{\mathbf{W}}_n^x)_{+}\left[\begin{array}{c}\overline{\mathbf{b}}_{x_{k-n}}^x\\ \vzero_{n\cdot m_u}\end{array}\right]+(\overline{\mathbf{W}}_n^x)_{-}\left[\begin{array}{c}\underline{\mathbf{b}}_{x_{k-n}}^x\\ \vzero_{n\cdot m_u}\end{array}\right]
\end{align*}
}%
where the weight and bias align with iterations in Theorem \ref{thm:recursivebound}. Thus  (\ref{eq:ueqstronger}) holds when $l=k$, thereby proving Theorem \ref{thm:recursivebound}.\qed
\end{pf}
\begin{cor}
Using the linear bound from Theorem \ref{thm:recursivebound} and the lower and upper bounds on \(\vu_0, \dots, \vu_{k-1}\), the reachable set \(\sR_k\) is derived via interval arithmetic \cite{liu-ia}.
\vspace{-0.1cm}
{
\small
\begin{equation}
\begin{aligned}
\label{eq:reachableset}
\sR_k &= \{\vx: \vx\geq_{\cdot} (\underline{\mathbf{W}}_k^x)_{+}\left[\begin{array}{@{}c@{}} \underline{\bm{x}}_{0} \\ \underline{\bm{u}}_{0}\\ \cdots \\ \underline{\bm{u}}_{k-1}\end{array}\right] + (\underline{\mathbf{W}}_k^x)_{-}\left[\begin{array}{@{}c@{}} \overline{\bm{x}}_{0} \\ \overline{\bm{u}}_{0}\\ \cdots \\ \overline{\bm{u}}_{k-1}\end{array}\right]+\underline{\mathbf{b}}_k^x,\\
& \vx\leq_{\cdot} (\overline{\mathbf{W}}_k^x)_{-}\left[\begin{array}{@{}c@{}} \underline{\bm{x}}_{0} \\ \underline{\bm{u}}_{0}\\ \cdots \\ \underline{\bm{u}}_{k-1}\end{array}\right] + (\overline{\mathbf{W}}_k^x)_{+}\left[\begin{array}{@{}c@{}} \overline{\bm{x}}_{0} \\ \overline{\bm{u}}_{0}\\ \cdots \\ \overline{\bm{u}}_{k-1}\end{array}\right]+\overline{\mathbf{b}}_k^x, \vx\in\sR^{m_x} \}
\end{aligned}
\end{equation}
}
where $\underline{\mathbf{W}}_k^x, \underline{\mathbf{b}}_k^x, \overline{\mathbf{W}}_k^x, \overline{\mathbf{b}}_k^x$ are weights and biases in (\ref{eq:ueq}). 
The symbols $\underline{(\cdot)}$ and $\overline{(\cdot)}$ denote the lower and upper bounds of \(\vx_0, \vu_0, \dots, \vu_{k-1}\), while $\geq_{\cdot}$ and $\leq_{\cdot}$ indicate element-wise comparison. 
\label{cor:reachable-set}
\end{cor}

We emphasize that the reachable set $\sR_k$ in Corollary \ref{cor:reachable-set} is dependent on $\underline{\vx}_0,\overline{\vx}_0, \underline{\vu}_0,\overline{\vu}_0,$ $\cdots, \underline{\vu}_{k-1},\overline{\vu}_{k-1}$, and $\sR_k$ can be referred to as a function of these variables. The function can be 
defined as $\sR_k = R_k^{\boldsymbol{\varepsilon}}(\vx_0,\vu_0,\cdots,\vu_{k-1})$, where $\boldsymbol{\varepsilon} = (\varepsilon_0, \varepsilon_1,\cdots,\varepsilon_{k-1})$, and $\vx_0 = \underline{\vx}_0,\overline{\vx}_0, \underline{\vu}_i = \vu_i-\varepsilon_i,\overline{\vu}_i = \vu_i+\varepsilon_i,i=0,1,\cdots,k-1$. The reachable set $\sR_k$ ensures that all perturbated states are within this set $\vx_k\in\sR_k$. 

From the reachability analysis, ensuring \(\sR_k \subseteq \sX\) in the nominal optimization problem guarantees constraint satisfaction for robust tracking. The nominal tracking problem at time step $k$ is then reformulated as follows:
\begin{prob}
\label{pb:robust-tracking-reformulation}
\begin{equation*}
    \begin{aligned}
        \text{min}& \sum_{i=k}^{k+N-1}((\hat{\vx}_{i}\!-\!\vr_{i})^\top\mQ (\hat{\vx}_{i}\!-\!\vr_{i})\!+\!\hat{\vu}_{i}^\top\mR \hat{\vu}_i)\\
        &+ (\hat{\vx}_{k+N}-\vr_{k+N})^\top\mQ_f (\hat{\vx}_{k+N}-\vr_{k+N})\\
        \text { s.t. } &  \hat{\vx}_{i+1} = \hat{\vx}_i + \vf(\hat{\vx}_i, \hat{\vu}_i) dt, \hat{\vx}_k = \vx_k\\
        & R_i^{\boldsymbol{\varepsilon}}(\hat{\vx}_k, \hat{\vu}_k,\cdots,\hat{\vu}_i)\subseteq\sX, R_{k+N}^{\boldsymbol{\varepsilon}}(\hat{\vx}_k, \hat{\vu}_k,\cdots,\hat{\vu}_{k+N})\subseteq\sX_f,\\
        & \hat{\vu}_i \in \sU\ominus\sW_i, i = k,k+1,\cdots,k+N-1
    \end{aligned}
\end{equation*}
where $\hat{\vx}_k = \vx_k$ is the given initial state and $\hat{\vu}_k,\hat{\vu}_{k+1},\cdots, $ $\hat{\vu}_{k+N-1}$ are descision variables. 
\end{prob}

After solving nominal optimization problem and obtain solutions $\hat{\vu}_k,\cdots,\hat{\vu}_{k+N-1}$, we apply control as $\vu_k = \hat{\vu}_k$ which guarantees constraint satisfaction in Problem~\ref{pb:robust-tracking} during control process with disturbance. 

% The key difference between the reachability-based formulation in Problem~\ref{pb:robust-tracking-reformulation} and the tube MPC problem in Problem~\ref{pb:robust-tracking-tube} lies in the treatment of state constraints. Tube MPC relies on a disturbance invariant set that is dependent on the linearized system dynamics matrices $\mA_k$ and $\mB_k$, which can lead to inaccuracies and over-conservatism, as previously noted. In contrast, the reachability-based problem uses time-dependent reachable sets as state constraints, which are computed directly from the neural network dynamics rather than linearized models. 
The key difference between Problem~\ref{pb:robust-tracking-tube} and Problem~\ref{pb:robust-tracking-reformulation} is in state constraint handling. Tube MPC relies on a disturbance invariant set based on linearized dynamics (\(\mA_k, \mB_k\)), leading to potential inaccuracies and conservatism. In contrast, the reachability-based approach leverages verification techniques to compute reachable sets directly from the neural network dynamics, avoiding approximations. The similarity between $\sZ_k$ and $\sR_k,\cdots,\sR_{k+N}$ is that they represent the distance from the nominal state and measure the level of perturbation. However, the time-dependent characteristic of the reachable set provides tighter bounds, reducing conservativeness and addressing tube MPC’s challenges, as shown in Section \ref{sec:experiment}.

% The relationship between the disturbance-invariant set $\sZ_k$ in Problem~\ref{pb:robust-tracking-tube} and the reachable set $\sR_i$ in Problem~\ref{pb:robust-tracking-reformulation} is straightforward. The disturbance invariant set $\sZ_k$ bounds the state error $\ve_i = \vx_i-\hat{\vx}_i$, whereas the reachable set $\sR_i$ bounds the perturbated state $\vx_i$. When applying the disturbance invariant set to the nominal state we obtain a reachable set of state for the tube MPC method, denoted as $R_i^{tube} = \hat{\vx}_i\oplus\sZ_k$. Comparing the conservativeness, i.e. the volume, of the disturbance invariant set in tube MPC and reachable set in our method is meaningful because both of them represent the distance from nominal state and measure the level of perturbation. However, the time-dependent nature of the reachable set in our method provides much tighter bounds, effectively reducing conservativeness, as demonstrated extensively in Section \ref{sec:experiment}. As a result, the challenges encountered in tube MPC can be addressed by employing the reachability-based method. 

% \begin{prop}
% Consider the existence of a feedback matrix $\mK$ that effectively bounds the reachable set $\sR_k$ as it evolves over an infinite horizon, ensuring that $\sR_k$ remains constrained as horizon $k$ increases. (To be completed)
% \end{prop}
\begin{rem}
Inspired by tube MPC, we could also control the growth of the reachable set through a feedback control as $\vu_k = \hat{\vu}_k+\mG_k(\vx_k-\hat{\vx}_k)$, where $\mG_k$ is the feedback matrix at time $k$. Using reachability tool, we can derive symbolic bound for $\vx_{k+1}$ as demonstrated in (\ref{eq:1layer})
\begin{align}
\vx_{k+1}\leq \overline{\mW}_{x_{k+1}}^f\vx_k+\overline{\mW}_{u_{k+1}}^f\vu_k+\overline{\vb}_{x_{k+1}}^x\\
\vx_{k+1}\geq \underline{\mW}_{x_{k+1}}^f\vx_k+\underline{\mW}_{u_{k+1}}^f\vu_k+\underline{\vb}_{x_{k+1}}^x
\end{align}
where $\left[\begin{array}{@{}c@{}c@{}}\overline{\mW}_{x_{k+1}}^f ;&\overline{\mW}_{u_{k+1}}^f\end{array}\right]\!=\!\overline{\mW}_{x_{k+1}}^x, \left[\begin{array}{@{}c@{}c@{}}\underline{\mW}_{x_{k+1}}^f ;&\underline{\mW}_{u_{k+1}}^f\end{array}\right]\!=\!\underline{\mW}_{x_{k+1}}^x$. We assume a simplified case where the difference of $\overline{\mW}_{x_{k+1}}^f-\underline{\mW}_{x_{k+1}}^f$ and $\overline{\mW}_{u_{k+1}}^f-\underline{\mW}_{u_{k+1}}^f$ are small and can be bounded as follows
% \vspace{-1cm}
\begin{align}
\|\overline{\mW}_{x_{k+1}}^f-\underline{\mW}_{x_{k+1}}^f\|_{\infty}\leq \gamma\\
\|\overline{\mW}_{u_{k+1}}^f-\underline{\mW}_{u_{k+1}}^f\|_{\infty}\leq \gamma
\end{align}

% Under this assumption, there exist a feedback matrix $\mG_k$ that simultaneously stabilizes both $\overline{\mW}_{x_{k+1}}^f + \overline{\mW}_{u_{k+1}}^f\mG_k$ and $\underline{\mW}_{x_{k+1}}^f + \underline{\mW}_{u_{k+1}}^f\mG_k$, which further leads to bounded reachable sets as $k$ increases over a long horizon. The simulation results in the DubinsCar environment, presented in Section \ref{sec:dubinscarexperiment}, confirm the existence of such a feedback matrix. Future work may explore more general feedback policies to further bound the reachable sets and how to obtain a forward invariant set in the presence of disturbances based on the bounded reachable sets. 

Under this assumption, a feedback matrix \(\mG_k\) can stabilize both \(\overline{\mW}_{\vx_{k+1}}^f + \overline{\mW}_{\vu_{k+1}}^f\mG_k\) and \(\underline{\mW}_{\vx_{k+1}}^f + \underline{\mW}_{\vu_{k+1}}^f\mG_k\) dynamics, ensuring bounded reachable sets over a long horizon. Simulations in the DubinsCar environment in Section~\ref{sec:dubinscarexperiment} confirm its existence. Future work may explore general feedback policies and forward invariant sets under disturbances.
\end{rem}

\subsection{Online control}
\label{reachabilitymethod}
% \begin{figure}[!t]
% \centering
%       \includegraphics[width=\columnwidth]{reachability-pipeline.pdf}
%       \vspace{-0.3cm}
%       \caption{Comparison between the proposed reachability-based method and the tube MPC method during online computation.}
%       \label{fig:pipeline}
% \end{figure}
Robust control is executed online by solving a nonlinear optimization at each step $k$, formulated as Problem~\ref{pb:robust-tracking-reformulation}. It minimizes a cost function while optimizing control inputs and reachable sets under reformulated state constraints for robustness guarantee. To efficiently solve this optimization, we use the IPOPT solver (Interior Point OPTimizer), as detailed in~\cite{ipopt}.

The optimization problem is inherently a Mixed Integer Programming (MIP) task. While NNDM dynamics are nonlinear, they can be decomposed into linear and mixed-integer constraints. Hidden layers in \(\vf\) are modeled as linear constraints, with integers encoding the ReLU activation layer~\cite{wei2022safe}. Future work will explore MIP solver implementation. The online execution process, contrasting with tube MPC, is shown in Fig. \ref{fig:nn-control-simu}.

\section{Numerical Study}
\label{sec:experiment}
% \begin{figure}[!b]
% \centering
%       \includegraphics[width=0.8\columnwidth]{css-scenarios2.pdf}
%       \caption{Illustration of Pendulum and DubinsCar scenarios.}
%       \label{fig:env-illu}
% \end{figure}

This section presents numerical examples demonstrating feedback control for bounding the reachable set over time and comparing tube MPC with reachability analysis in terms of conservativeness and control cost for robust tracking. We first outline the problem settings in the Pendulum and DubinsCar environments from \texttt{RobotZoo.jl}: \url{https://github.com/RoboticExplorationLab/RobotZoo.jl}, as shown in Fig.~\ref{fig:nn-control-simu}, then evaluate different approaches.

\subsection{Simulation set up}
% We utilize neural network-approximated dynamics for the Pendulum and DubinsCar environments, trained via a supervised learning approach with mean square error (MSE) as the loss function. Well-fitted neural networks are treated as the true dynamics in (\ref{eq:nndm}), with a discrete time interval of $dt = 0.1$. Perturbations on the control input along the trajectory are randomly sampled within a constant $\vl_{\infty}$ bound, implying that $\sW_0 = \sW_1 = \cdots = \sW_i = \cdots = \sW = \{\vw \in \sR | \|\vw\|_{\infty} \leq \varepsilon \}$. For both scenarios, we consider a 1-step look-ahead robust tracking problem without loss of generality. Extending this to a multi-step look-ahead will be addressed in future work.
% These examples are implemented in Julia, and nominal tracking problems are solved using IPOPT \citep{ipopt}.

% The comparison is made between our proposed method and the conventional tube MPC method, introduced in Section \ref{pl:tubempc} and Section \ref{reachabilitymethod}, respectively. In the following figures and tables, the label \emph{Baseline} refers to the tube-based approach, while \emph{Ours} denotes the reachability-based robust MPC for NNDMs.

We use neural network-approximated dynamics for the Pendulum and DubinsCar environments, trained via supervised learning with MSE loss. The fitted networks serve as true dynamics in (\ref{eq:nndm}) with a discrete time step $dt = 0.1$. Control input perturbations are randomly sampled within a constant $\ell_{\infty}$ bound, i.e., \( \sW_0 = \cdots = \sW_i = \cdots =\sW = \{\vw \in \sR \mid \|\vw\|_{\infty} \leq \varepsilon \} \). We consider a 1-step look-ahead optimization problem, i.e. $N=1$, with multi-step extensions left for future work.

Implemented in Julia, the nominal tracking problems are solved using IPOPT. We compare the effectiveness of our reachability-based method (\emph{Ours}) with the tube MPC method (\emph{Baseline}). Tube MPC is implemented according to Section \ref{pl:tubempc}, and the feedback matrix $\mK_k$ can be calculated using the \texttt{dlqr} function. 
% Given $\mA_k = \begin{bmatrix} 1.00 & 0 & 0.12\\ 0 & 1.0 & 0.05 \\ 0.08 & -0.05 & 2.0 \end{bmatrix},\mB_k =\begin{bmatrix} 0.09 & 0.05 \\ 0.03 & -0.03 \\ -0.09 & 2.22 \end{bmatrix},\mQ = \mI_3, \mR =0.01\mI_2$, we can use dlqr to compute feedback matrix $\mK_k = \begin{bmatrix} 6.11 & 1.67 & -0.08 \\ 0.17 & 0.47 & 0.94 \end{bmatrix}$.

\subsection{Robust tracking problem design}
In the Pendulum environment, the state and action are $\vx = [\theta, \dot{\theta}]^\top, \vu = \tau$, where $\theta$ and $\dot{\theta}$ denote angle and angular velocity, and $\tau$ is the applied torque, as shown in Fig.~\ref{fig:nn-control-simu}. The constraints are $\sX = \{\vx\in\sR^2|\ [-\pi,\pi]\times[-10,10]\}, \sU = \{u\in\sR|\ |u|\leq 10\}, \sW = \{\vw\in\sR|\ \|\vw\|_{\infty}\leq \varepsilon\}$. The reference trajectory follows a constant angular velocity: $\theta_r = 0.12t, \dot{\theta}_r = 0.12$. The cost matrices are $\mQ = \mI_2, \mR = 0.01$. 

In DubinsCar, the state and action are $\vx\!=\![p_x, p_y, \alpha]^\top, \vu\!=\![v,\phi]$, where $p_x, p_y$ are positions, $\alpha$ is the orientation, and $v, \phi$ are the velocity and angular velocity, respectively, detailed in Fig.~\ref{fig:nn-control-simu}. The constraints are $|p_x|,|p_y|\leq 1$, $|\alpha|\leq\pi$, $|v|,|\phi|\leq 5$ and 
 $\sW = \{\vw\in\sR| \|\vw\|_{\infty}\leq \varepsilon\}$. In this scenario, a planner assists in reference tracking. The state at time $k$ is $\vx_k\!=\![p_x^k, p_y^k, \alpha^k]^\top$, with the target state $\vx_r = [p_x^r,p_y^r,0]^\top$. The planning state $\vx_{p}^k\!=\![p_{x,p}^k,p_{y,p}^k,\alpha_p^k]^\top$ is generated at each step $k$, where $\alpha_p^k\!=\!\arctan\left(\frac{p_y^r-p_y^k}{p_x^r-p_x^k}\right)$ and $[p_{x,p}^k,p_{y,p}^k]^\top$ is a segment point along the line from $[p_x^k,p_y^k]^\top$ to $[p_x^r,p_y^r]^\top$, with a projection limit of 0.08. The optimization problem treats $\vx_p^k$ as a reference, with cost matrices $\mQ = \mI_3, \mR = 0.01\mI_2$. 
 
\subsection{Results and discussion}
\subsubsection{Pendulum}
\begin{figure}[!t]
\centering
      \includegraphics[width=0.6\columnwidth]{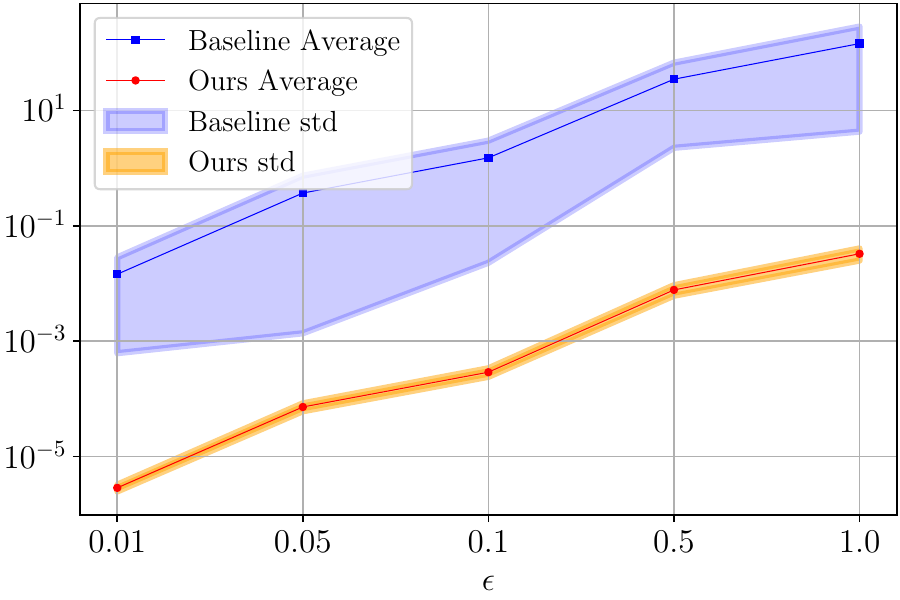}
      \vspace{-0.5cm}
      \caption{Comparison of the average and variance of the set area between the tube MPC and reachability-based methods with varying levels of perturbation.}
      \label{fig:pen-con-static}
\end{figure}

\begin{figure}[!t]
\vspace{-0.3cm}
\centering
      \includegraphics[width=0.45\columnwidth]{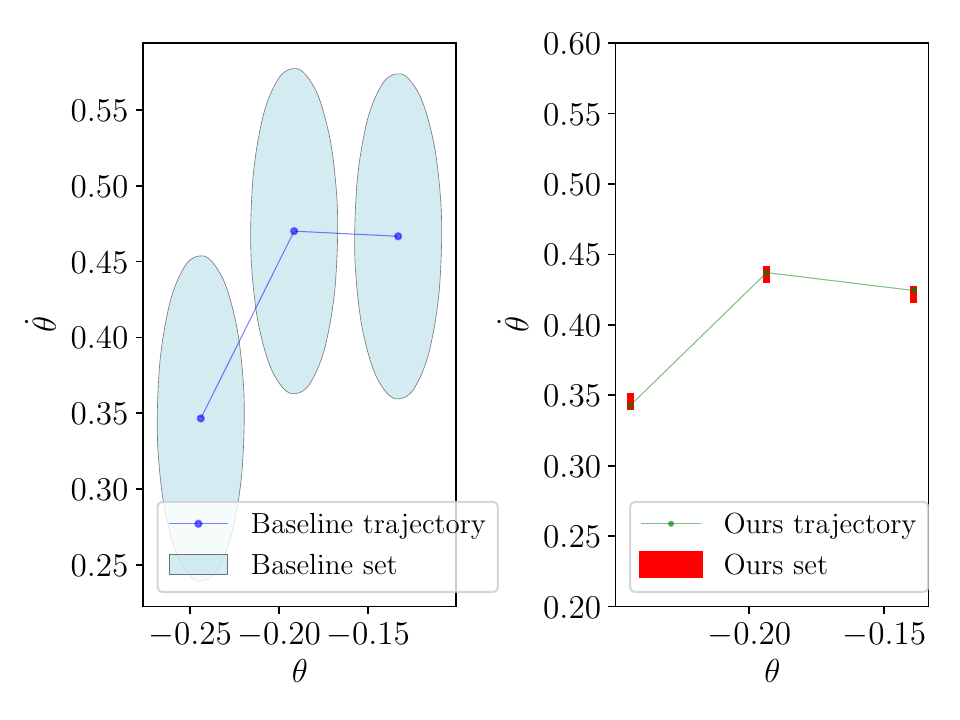}
       \includegraphics[width=0.45\columnwidth]{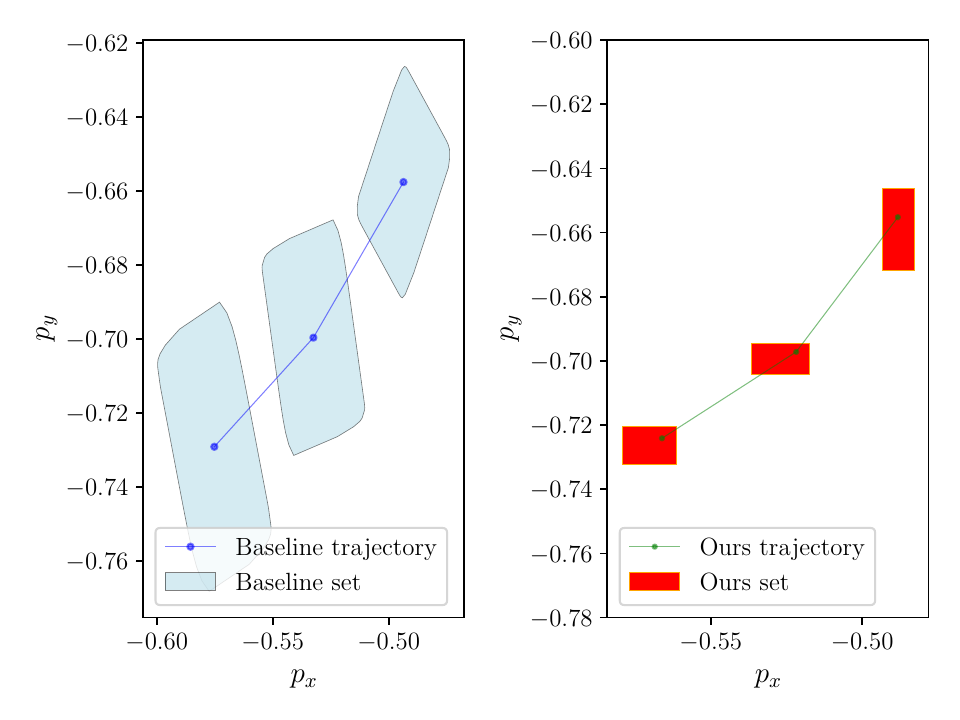}
       \vspace{-0.4cm}
      \caption{Comparison of the disturbance invariant set and reachable set between the tube MPC method and reachability-based method. Left: Pendulum, $\varepsilon = 0.01$; Right: DubinsCar, $\varepsilon = 0.1$.}
      \label{fig:set}
\end{figure}

We randomly sample 50 initial states from $\{[\theta, \dot{\theta}]^\top |\ |\theta| \leq \pi/2, \dot{\theta} = 0\}$ and apply control for 50 steps for each instance. The left plot in Fig.~\ref{fig:set} depicts the disturbance invariant set and the reachable set along trajectories for $\varepsilon = 0.01$ in robust tracking. Additionally, Fig. \ref{fig:pen-con-static} compares the conservativeness of the reachability-based and tube MPC methods by evaluating the areas of the reachable sets and disturbance sets. Solid lines represent the average set area per step, while ribbons indicate standard deviations. As shown, the reachability-based method results in a significantly smaller average areas and a narrower variation band, reflecting consistent stability and reduced conservativeness throughout the trajectory.

\begin{figure}[!t]
\vspace{-0.3cm}
    \centering 
    \includegraphics[width=0.7\columnwidth]{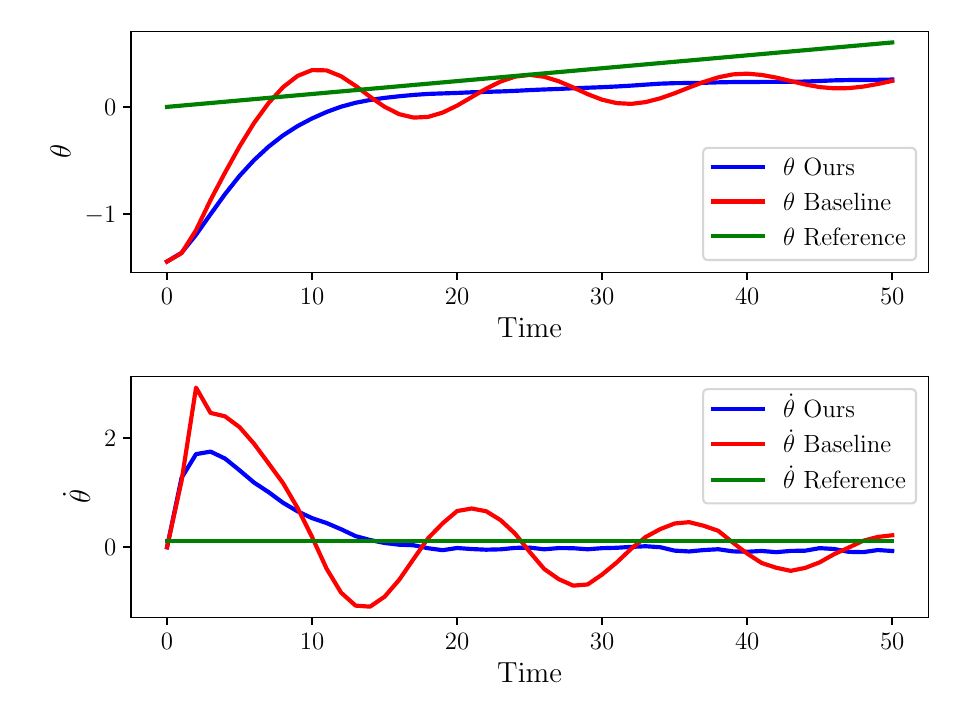}
      % \resizebox{0.5\columnwidth}{!}{\input{control_traj_Pendulum_0.1.tex}}
      
      \vspace{-0.5cm}
      \caption{Trajectory comparison for $\varepsilon = 0.1$. Constant tracking error may result from a short tracking horizon and persistent control input disturbance.}
      \label{fig:pen-traj}
\end{figure}

Fig. \ref{fig:pen-traj} depicts trajectories under perturbation $\varepsilon= 0.1$. Reachability-based method achieves more accurate and stable tracking with smaller state errors, effectively following the lifting pendulum without oscillations. 
In contrast, tube MPC method exhibits continuous oscillations, with $\theta$ and $\dot{\theta}$ varying sinusoidally. This instability likely stems from the conservativeness of the disturbance invariant set. 
As a result, the reachability-based method provides a tighter reachable set, enabling more aggressive torque inputs to counteract gravity, ensuring smooth tracking. Tube MPC’s conservative set leads to insufficient control adjustments, causing persistent oscillations and poor tracking.
% Our method provides a tight reachable set, which allows the optimization problem to compute more aggressive torque inputs to counteract the gravitational forces acting on the pendulum, resulting in a slow but stable trajectory tracking. On the other hand, the disturbance invariant set provided by tube MPC method is conservative, causing the solutions of optimization problem too insufficient to counteract gravity effectively. As a result, the pendulum repeatedly oscillates, failing to adequately track the reference state. 

\subsubsection{DubinsCar}
\label{sec:dubinscarexperiment}
\vspace{-0.2cm}
\begin{figure}[!t]
\centering
      \includegraphics[width=0.7\columnwidth]{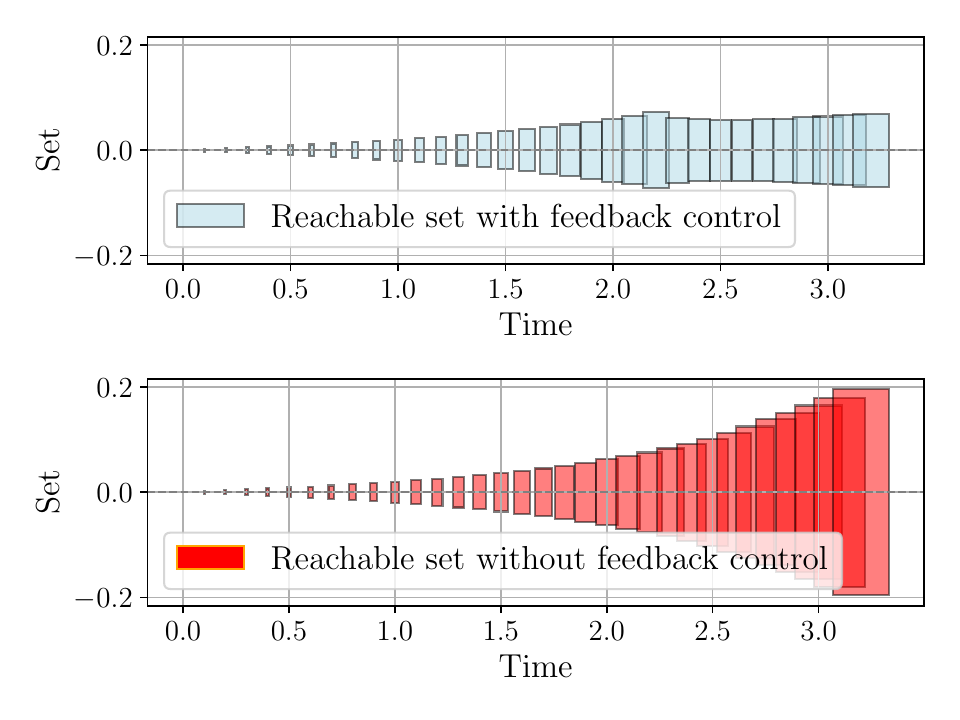}
      \vspace{-0.5cm}
      \caption{
      Reachable sets with and without feedback over time for $\varepsilon = 0.01$. The center remains at $y = 0$, with the $x$-axis representing simulation time ($dt = 0.1$). While the DubinsCar model has three state dimensions, only the first two are shown for clarity, omitting the orientation angle $\theta\in[-\pi, \pi]$.}
      \label{fig:feedbackreachability}
\end{figure}
First, we sample a initial state from the set $\{[p_x, p_y, \alpha]^\top|\ |p_x|,|p_y|\leq 1, \alpha=0\}$ and apply control for 35 steps. Fig. \ref{fig:feedbackreachability} compares the evolution of reachable sets with and without feedback. With feedback, the reachable set gradually expands but remains bounded, whereas without feedback, it grows exponentially. 
% Also we show statistic graph of average and std of reachable sets volume over 20 control instances, each control for 30 steps in \Cref{fig:static-reachable}. It shows that as control step increases, the volume of reachable set with feedback is consistently lower than no feedback control, with lower std. 

 Next, we randomly sample 20 initial and target states from $\{[p_x, p_y, \alpha]^\top|\ |p_x|,|p_y|\leq 1, \alpha=0\}$ and apply control for 20 steps. The right plot of Fig. \ref{fig:set} illustrates the disturbance invariant set and reachable set for $\varepsilon = 0.1$. Observation indicates that the tube MPC method remains more conservative than our method. 

 \begin{figure}[!t]
 \vspace{-0.3cm}
\centering
      \includegraphics[width=0.9\columnwidth]{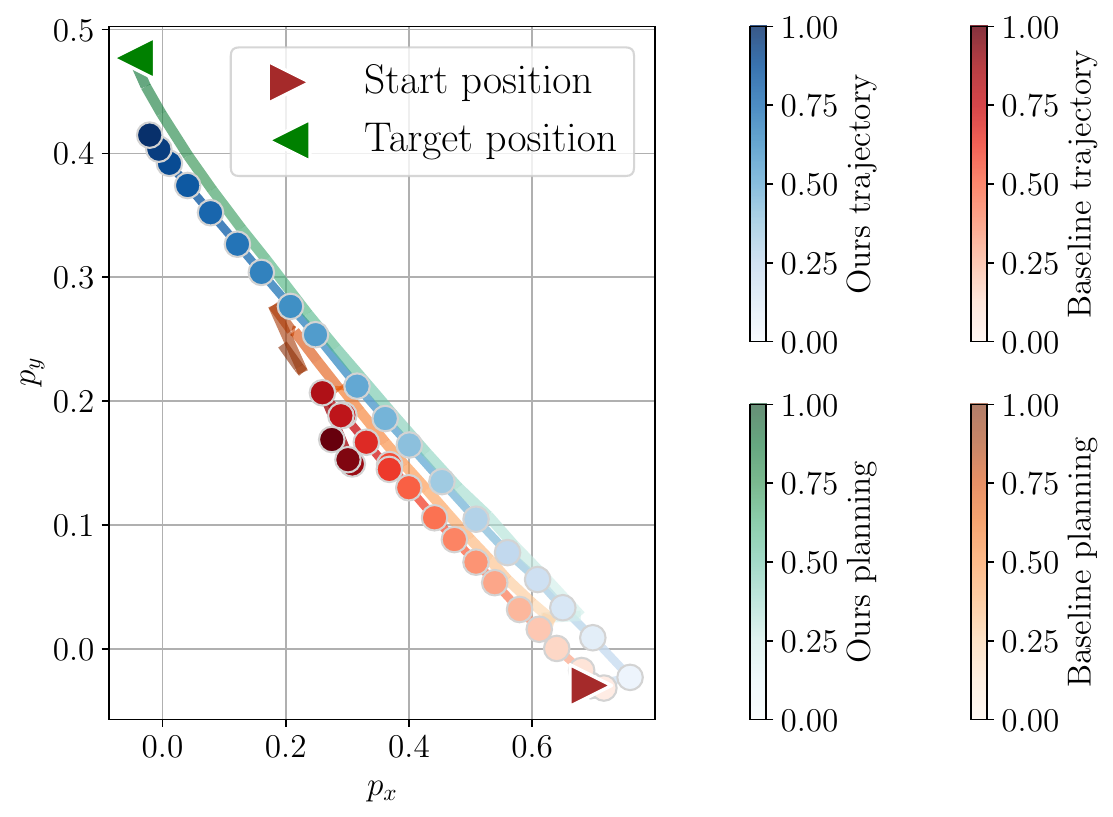}
      \vspace{-0.4cm}
      \caption{Comparison of trajectories from different methods when $\varepsilon=0.1$. The start $(p_x^i,p_y^i)$ and target $(p_x^r,p_y^r)$ positions are the first two dimensions of $\vx_0$ and $\vx_r$. The color bar shows the car’s movement over time, with darker shades for later steps.}
      \label{fig:dub-traj}
\end{figure}

  Fig. \ref{fig:dub-traj} presents an example trajectory when $\varepsilon\!=\!0.1$. In this figure, our method consistently follows the planned path and reaches the target position effectively, while tube MPC trajectory hovers near the start before slowly approaches the target. The behavior of tube MPC may stem from an overly conservative disturbance invariant set or inaccuracies in linearized dynamics, leading to suboptimal control.

\begin{table}[!t]
\caption{Average per-step cost and error across methods under varying perturbations in Pendulum and DubinsCar.}
\label{tab:cost}
\begin{tabular}{l|llllll}
\hline
Pendulum          & \multicolumn{3}{l}{Ave per-step cost}         & \multicolumn{3}{l}{Ave per-step err}          \\ \hline
$\varepsilon$ & 1.0  & 0.1  & 0.01 & 1.0  & 0.1  & 0.01 \\ \hline
Baseline    & 0.59 & 0.15 & 0.11 & 1.58 & 0.62 & 0.35 \\ \hline
Reachability-based & \textbf{0.18} & \textbf{0.13} & \textbf{0.11} & \textbf{0.58} & \textbf{0.38} & \textbf{0.34} \\ \hline
\end{tabular}
\hfill
\vspace{0.3cm}
\begin{tabular}{l|llllll}
\hline
DubinsCar          & \multicolumn{3}{l}{Ave per-step cost}         & \multicolumn{3}{l}{Ave per-step err}          \\ \hline
$\varepsilon$ & 0.1  & 0.01 & 0.001 & 0.1  & 0.01 & 0.001 \\ \hline
Baseline    & 0.53 & 0.45 & 0.68  & 0.30 & 0.28 & 0.27  \\ \hline
Reachability-based & \textbf{0.32} & \textbf{0.41} & \textbf{0.64} & \textbf{0.23} & \textbf{0.21} & \textbf{0.25} \\ \hline
\end{tabular}
\end{table}

% \usepackage{multirow}

\begin{table}[!t]
\vspace{-0.2cm}
\caption{Computation time comparison for disturbance invariant and reachable sets, along with IPOPT solving time, across methods and scenarios for $\varepsilon=0.1$.}
\label{tab:time}
\begin{tabular}{llllll}
\hline
\multicolumn{2}{l|}{\multirow{2}{*}{\begin{tabular}[c]{@{}l@{}}Time(s)\\ $\varepsilon = 0.1$\end{tabular}}} &
  \multicolumn{2}{l}{Set} &
  \multicolumn{2}{l}{IPOPT} \\ \cline{3-6} 
\multicolumn{2}{l|}{}                                   & Baseline & Ours & Baseline & Ours \\ \hline
\multirow{2}{*}{Pendulum}  & \multicolumn{1}{l|}{Average} & 0.009    & \textbf{0.004}     & 0.222    & \textbf{0.018}     \\ \cline{3-6} 
                           & \multicolumn{1}{l|}{Std}  & 0.004    & \textbf{0.002}     & 0.144    & \textbf{0.014}     \\ \hline
\multirow{2}{*}{DubinsCar} & \multicolumn{1}{l|}{Average} & 0.744    & \textbf{0.007}     & 4.830    & \textbf{2.762}     \\ \cline{3-6} 
                           & \multicolumn{1}{l|}{Std}  & 0.051    & \textbf{0.002}     & 3.548    & \textbf{2.688}     \\ \hline
\end{tabular}
\end{table}

Additionally, Table \ref{tab:cost} compares average per-step control costs and average per-step tracking error $\ve_k = \|\vx_k-\vr_k\|_{2}$, across  methods under varying disturbance bounds. The results demonstrate that reachability-based method largely reduces state deviation and control cost, outperforming tube MPC. Table \ref{tab:time} highlights the computational advantage of reachability-based in reachable set computation and IPOPT solving, compared to tube MPC's disturbance invariant set computation and optimization. The results are based on a 1-step look-ahead and further experiments are needed for multi-step scenarios. 

\begin{figure}[!t]
\centering
      \includegraphics[width=0.9\columnwidth]{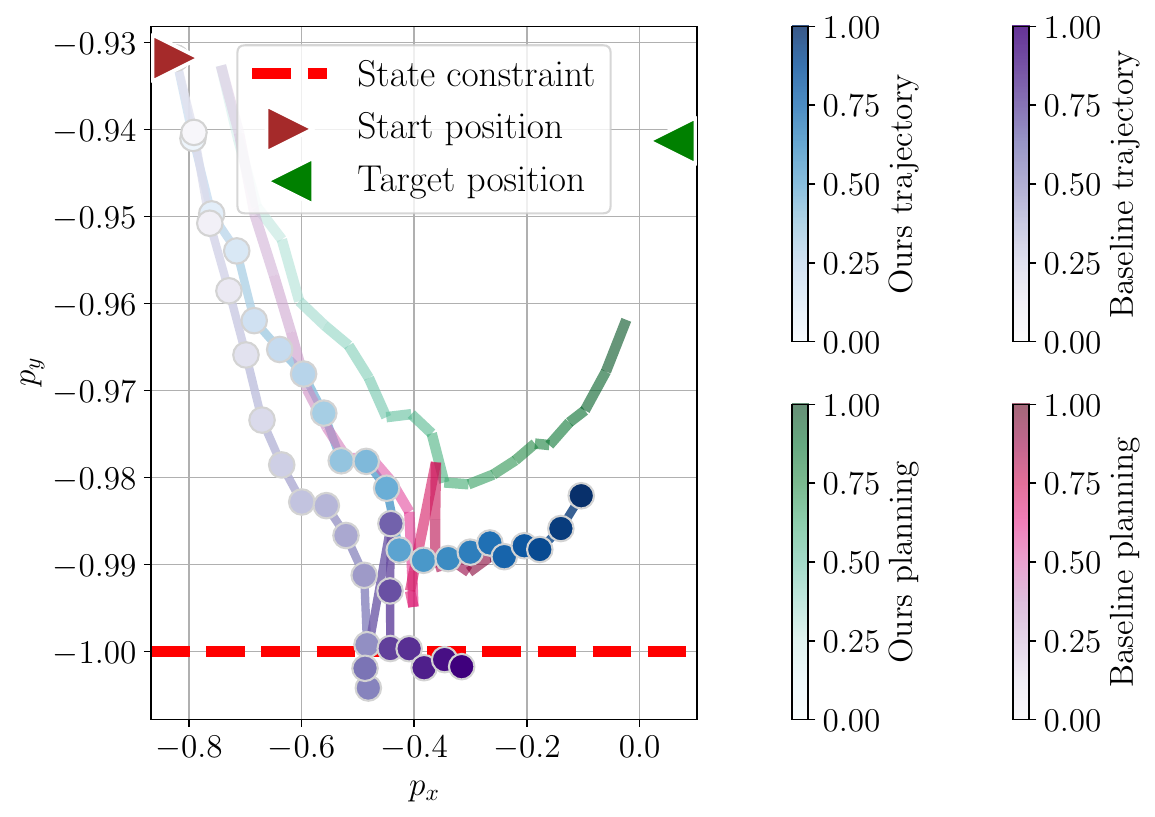}
      \caption{Comparison of trajectories from different methods when $\varepsilon=0.1$. The color bar shows the car’s movement over time, with darker shades for later steps.}
      \label{fig:dub-constraint-traj}
\end{figure}

Furthermore, we compare the reachability-based method with a naive approach that ignores disturbances when optimizing but considers them in control. As shown in Fig. \ref{fig:nn-control-simu}, tracking tasks are set near the constraint boundaries, while expecting trajectories to stay within constraints without violations. Table \ref{tab:dub-constraint-tab} and Fig. \ref{fig:dub-constraint-traj} show that the reachability-based outperforms the naive one, which applies aggressive control, causing violations and higher costs. In contrast, the reachability-based method successfully balances conservativeness and robustness.

\begin{table}[!t]
\caption{Average per-step cost and error across methods under varying perturbations in DubinsCar, with states near constraints.}
\label{tab:dub-constraint-tab}
\begin{tabular}{l|llllll}
\hline
DubinsCar     & \multicolumn{3}{l}{Ave per-step cost} & \multicolumn{3}{l}{Ave per-step err} \\ \hline
$\varepsilon$ & 1.0         & 0.1        & 0.01       & 1.0              & 0.1     & 0.01    \\ \hline
Naive      & 1.023      & 0.39      & 0.35      & \textbf{0.09}   & 0.06   & 0.05   \\ \hline
Reachability-based & \textbf{0.99} & \textbf{0.36} & \textbf{0.3} & 0.15 & \textbf{0.05} & \textbf{0.05} \\ \hline
\end{tabular}
\end{table}

%%%%%%%%%%%%%%%%%%%%%%%%%%%%%%%%%%%%%%%%%%%%
\section{Conclusion}

This paper presents a reachability-based robust control method for discrete-time neural network dynamic models (NNDM) under bounded input perturbations. We introduce a specialized $k$-step reachability tool for NNDM and integrate it into a robust tracking framework to ensure constraint satisfaction. Simulations demonstrate the method’s effectiveness in reducing conservatism and control cost compared to tube MPC. Future work will explore mixed-integer programming for exact solutions \citep{wei2022safe}, Bernstein polynomial approximations for faster computation \citep{hu2024real}, and forward invariant sets for persistent feasibility.
\bibliography{root}

\end{document}

%% file: math_commands.tex
\usepackage{amsmath,amsfonts,bm}

\def\eqref#1{equation~\ref{#1}}
\def\1{\bm{1}}

\def\vzero{{\bm{0}}}

\def\va{{\bm{a}}}
\def\vb{{\bm{b}}}

\def\ve{{\bm{e}}}
\def\vf{{\bm{f}}}

\def\vh{{\bm{h}}}

\def\vr{{\bm{r}}}

\def\vu{{\bm{u}}}

\def\vw{{\bm{w}}}
\def\vx{{\bm{x}}}

\def\vz{{\bm{z}}}

\def\evb{{b}}

\def\mA{{\bm{A}}}
\def\mB{{\bm{B}}}

\def\mG{{\bm{G}}}
\def\mH{{\bm{H}}}
\def\mI{{\bm{I}}}

\def\mK{{\bm{K}}}

\def\mM{{\bm{M}}}

\def\mQ{{\bm{Q}}}
\def\mR{{\bm{R}}}

\def\mW{{\bm{W}}}

\DeclareMathAlphabet{\mathsfit}{\encodingdefault}{\sfdefault}{m}{sl}
\SetMathAlphabet{\mathsfit}{bold}{\encodingdefault}{\sfdefault}{bx}{n}

\def\sA{{\mathbb{A}}}
\def\sB{{\mathbb{B}}}

\def\sR{{\mathbb{R}}}

\def\sU{{\mathbb{U}}}

\def\sW{{\mathbb{W}}}
\def\sX{{\mathbb{X}}}

\def\sZ{{\mathbb{Z}}}

\def\emW{{W}}